# Wireless ad-hoc networks: Strategies and Scaling laws for the fixed SNR regime [*]

Shuchin Aeron    Venkatesh Saligrama
Department of Electrical and Computer Engineering
Boston University, MA 02215
Email {shuchin,srv}@bu.edu

**Abstract**

This paper deals with throughput scaling laws for random ad-hoc wireless networks in a rich scattering environment. We develop schemes to optimize the ratio, $\rho(n)$ of achievable network sum capacity to the sum of the point-to-point capacities of source-destinations pairs operating in isolation. For fixed SNR networks, i.e., where the worst case SNR over the source-destination pairs is fixed independent of $n$, we show that collaborative strategies yield a scaling law of $\rho(n) = \mathcal{O}(\frac{1}{n^{1/3}})$ in contrast to multi-hop strategies which yield a scaling law of $\rho(n) = \mathcal{O}(\frac{1}{\sqrt{n}})$. While, networks where worst case SNR goes to zero, do not preclude the possibility of collaboration, multi-hop strategies achieve optimal throughput. The plausible reason is that the gains due to collaboration cannot offset the effect of vanishing receive SNR. This suggests that for fixed SNR networks, a network designer should look for network protocols that exploit collaboration. The fact that most current networks operate in a fixed SNR interference limited environment provides further motivation for considering this regime.

## 1 Introduction

We consider a network of $n$ source-destination (S-D) pairs where the sources and destinations are deployed uniformly in a region of space. Each source can communicate to the intended destination over a wireless channel. The wireless channel is assumed to undergo the usual attenuation and has fading. Owing to the presence of multiple S-D pairs, it becomes necessary to share the channel for simultaneously serving all the S-D pairs. The problem is to determine maximum achievable throughput for large wireless networks. Our paper deals with this problem for fixed SNR rich scattering environments.

The general problem of addressing how the network throughput scales as a function of the number of S-D pairs has been a subject of intensive research [1, 2, 3, 4, 5, 6, 7, 8, 9, 10, 11, 12, 13]. The seminal work of [1] introduced the problem of capacity of wireless networks and this problem was studied further from an information-theoretic viewpoint in [2, 3, 4, 5]. In [7] capacity of wireless networks where nodes were distributed on a line was considered. In [8, 9] networks with heterogenous nature that involve both wireless and wired infrastructure was studied motivated by practical scenarios. In [10, 11] wireless networks with mobile nodes were considered. In [12] capacity of wireless networks with multiple antennas at each node is considered motivated by the large rate gains achievable by the multiple antenna systems. Recently capacity of arbitrary wireless

[*]This research was supported by the ONR Young Investigator Program and Presidential Early Career Award (PECASE) N00014-02-100362, NSF CAREER award ECS 0449194, and NSF Grant CCF 0430983 and CNS-0435353

networks, was studied in [13] based on a network flow graph abstraction of the protocol model of [1]. In [14] percolation theory arguments were used to determine the capacity of wireless networks.

Unlike traditional networks, geometry and locations of the S-D pairs play an important role in this context. For example if each source is sufficiently close to its destination and there is sufficient separation between different S-D pairs, each S-D pair can then communicate with each other at a fixed positive rate independent of $n$, for sufficiently large attenuation in the media. If however the source and destination locations are arbitrary then in order for a source to communicate successfully to the intended destination, then each S-D pair must share the channel with other such pairs, which in general leads to reduction of capacity.

The problem in its fairly general form was studied in the seminal paper [1], where the sources and destinations were uniformly placed in a 2-d region of space. Under the particular type of network protocol, namely multi-hop protocol, they showed that the throughput of such a network scales as $\mathcal{O}(\frac{1}{\sqrt{n}})$ bits/sec/S-D pair. From an informational perspective, multi-hop protocol is an interference avoidance protocol where direct communication between any two nodes $A$ and $B$ precludes transmission from any source within a communication radius of destination $B$. Consequently, this results in a trade off between the number of S-D pairs suppressed and the traffic created as a result of intermediate relaying and it turns out that a multi-hop strategy with minimum radius of communication is optimal.

This has led to consideration of optimal schemes from an information theoretic perspective, which allow for multiple simultaneous transmissions and utilize interference cancellation techniques. In this context two types of network scenarios have been considered: (1) Fixed area networks: where a fixed area is populated with more and more nodes leading to a dense network. (2) Extended area networks: where a constant distance is maintained between any two nodes leading to increase in the geographical expanse of the network with the number of nodes. For both types of scenarios upper bounds to the network throughput are considered from an information theoretic perspective in [5, 4, 2] and more recently in [6]. Under fixed power $P$ at each node it is shown that for extended networks the information theoretic upper bound scales as $\mathcal{O}(\frac{\log n}{\sqrt{n}})$ bits/sec/S-D pair. This is not far from what is achievable via multi-hop strategies. In contrast it is shown in [5] the information theoretic upper bound for the fixed area network (with power constraint $P$ at each node) is $\mathcal{O}(1)$ bits/sec/S-D pair, and they called for strategies to close this gap. This paper presents first results towards closing this gap.

The relatively small gap between upper and lower bounds for extended area networks can be ascribed to the following reasoning. Given that the network size is increasing, the distance between most S-D pairs increase with $n$. If we preclude any other S-D communication, the capacity between a S-D pair goes to zero for a fixed power $P$. Use of collaborative strategies in this case may help reduce the impact of multi-user interference cancelation. Nevertheless, the throughput is dominated by the vanishing SNR, i.e. the power attenuation offsets the gains due to collaboration. On the other hand multi-hop strategies provide a mechanism to improve the throughput. In contrast for fixed area networks the receive SNR between any S-D pair is non-vanishing and is independent of $n$ due to a fixed worst-case SNR. In this case multi-user interference is the principal throughput limiting factor. Such wireless networks with fixed worst case SNR over all S-D pairs is the subject of present paper.

We introduce a fundamental unitless network metric in this context, the ratio, $\rho(n)$, of the achievable sum capacity of the network of S-D pairs to the sum capacity of S-D pairs operating in isolation. This metric captures the effect of interference in a network by normalizing out the effect of point-to-point capacity of any S-D pair. Also, normalizing the point-to-point capacity in a fixed SNR network removes any differences between extended area and fixed area networks. To illustrate these points further in the following example.

*Example* 1.1. Suppose the power attenuates as, $d^{-4}$, where $d$ is transmission distance. The point-to-point capacity between a typical S-D pair in an $n$-node extended area network scales as $n^{-2}$ bits/sec if the two nodes were to operate in isolation, while a rate of $n^{-0.5}$ is achievable through multi-hop communications as indicated by results in the literature [4, 1, 5, 3, 2]. Thus, the ratio $\rho(n) = n\sqrt{n}$. In contrast if each node has sufficient power to maintain a fixed worst case SNR then, the ratio is $\rho(n) = \frac{1}{\sqrt{n}}$ for a multi-hop protocol. This leads to the plausible argument that attenuation is the dominant factor in the reduced throughput capacity of an extended area network and the network provides much benefit in improving the throughput capacity. Moreover, it is also apparent from the example that for a fixed SNR network interference is the principal throughput governing factor since attenuation does not play a significant role. The fact that most current networks operate in a fixed SNR interference limited environment provides further motivation for considering this regime.

In this paper we show that under the fixed SNR regime the network sum capacity scales as $\mathcal{O}(n^{2/3})$ bits/sec. This is a significant improvement over the $\mathcal{O}(\sqrt{n})$ bits/sec result of [1]. Consequently the network metric $\rho(n)$ scales as $\mathcal{O}(\frac{1}{n^{1/3}})$ in contrast to $\mathcal{O}(\frac{1}{\sqrt{n}})$ obtainable with multi-hop protocols.

The paper is organized as follows. In section 2 we will describe the problem set-up with precise notion of fixed SNR regime which is the subject of the present paper. In section 4 we introduce notation to precisely define the collaborative scheme proposed in the paper. In section 5 we will present the outline of the collaborative scheme and comment on the schematic aspects of the scheme. Subsequently in section 6 we will provide the proof of the main result. Finally we conclude in section 7.

## 2 Problem Set-up

In this section we present the communication model and the problem setup. We consider a square region $\mathcal{Z}$ with $n$ nodes on a regular grid with coordinates $(j\rho_{\min}, k\rho_{\min})$, $j, k = 1, 2, \ldots, \sqrt{n}$, where $\rho_{\min}$ is the minimum distance between the nodes. The channel is a standard frequency non-selective channel with independent fades between any two nodes. Specifically, the communication model from node $p$ to node $q$ is:

$$Y_q = \frac{\sqrt{P}}{d_{qp}^{\alpha/2}} h_{qp} X_p + N_q \tag{1}$$

where, $P$ is node $p$'s transmit power, $X_p$ is the transmitted symbol, $d_{qp}$ is the distance separating the two nodes, and $\alpha$ is the attenuation coefficient. $h_{qp}$ is the fading gain from transmitter node $p$ to receiver node $q$. We assume that $h_{qp}$ is i.i.d. $\sim \mathbb{CN}(0,1)$ for all $p, q$, where $\mathbb{CN}(0, \mathbf{R})$ means complex Gaussian vector with zero mean and covariance $\mathbf{R}$. The receiver noise $N_q$ is independent and is AWGN with variance $N_0$.

**Scale invariant communication model:** We propose a scale invariant communication model to account for changing number of nodes, and network topology. Our subsequent analysis of the network will be based on this model.

The communication model of equation (1) is a standard model for rich scattering environments [15, 12, 16, 17]. We enforce the fixed-SNR model in the network as follows: Normalize with respect to the maximum admissible SNR to the distant nodes and scale the SNR in proportion to the distance to the neighboring nodes, i.e.,

$$Y_q = \sqrt{SNR} \left(\frac{d_{\max}}{d_{qp}}\right)^{\alpha/2} h_{qp} X_p + N_q, \ E(X_p^2) \leq 1 \tag{2}$$

where, $N_q$ is AWGN noise with noise variance equal to one and $SNR \leq SNR_0$. $SNR_0$ is the worst-case signal-to-noise between any two S-D pairs; $d_{\max}$ is the maximum of the distances between S-D pairs. $X_p$ is the symbol transmitted by node $p$. The symbol power is bounded by one to be consistent with the model of Equation 1. The above communication model is a *scale invariant model* since the SNR between two maximally distant nodes is held constant irrespective of the network size. Also, the minimum distance $\rho_{\min}$ is no longer a factor in the new model.

The model of Equation (1) and Equation (2) are worst case SNR equivalent. To see this, suppose that each node has power $P$. Then by definition $SNR_0 = \frac{P}{d_{\max}^\alpha N_0}$ based on Equation (2). The $SNR$ at node $q$ from node $p$ is given by $\frac{P}{d_{pq}^\alpha N_0}$. The ratio $\frac{SNR_q}{SNR_0}$ is then equal to $\frac{d_{\max}^\alpha}{d_{pq}^\alpha}$ which is consistent with Equation (2).

To form S-D pairs we partition the square region into equal rectangular transmitter and receiver regions, $\mathcal{Z}_1$, $\mathcal{Z}_2$ with $n/2$ nodes in each region. For each node in $\mathcal{Z}_1$ we randomly select a destination in $\mathcal{Z}_2$ such that for any two nodes in $\mathcal{Z}_1$ the corresponding destinations are different and for any two nodes in $\mathcal{Z}_2$ the corresponding source nodes are different.

The objective is to determine how the capacity per S-D pair scales with the number of nodes, $n$, in the network. We formalize this objective as follows: compute the ratio between the sum-capacity over all S-D pairs in the network and the sum of the point-to-point capacity for each S-D pair, i.e.,

$$\rho(n) = \sup \frac{\sum_{s,d} C_n^{sd}(P)}{\sum_{s,d} C_0^{sd}(P)} \tag{3}$$

where, $C_n^{sd}(P)$, $C_0^{sd}(P)$ are the capacities for the S-D pair $(s,d)$ while operating in a network and in isolation (with no multi-user interference) respectively. The supremum in Equation (3) is over all the communication strategies, which are all the admissible coding and collaboration strategies under the communication model.

We briefly comment on our choice of S-D pairs. While it is conventional to choose the S-D pairs randomly in the entire region, our preference for the partitioned model is to avoid unnecessary technical details and maintain simplicity of exposition. Still, it is worth pointing out that the typical S-D distance in the partitioned model is of the same order i.e. $\mathcal{O}(\sqrt{n})$ as that obtained by choosing S-D pairs randomly, c.f. [1]. Therefore, the transport capacity metric in bit-meters/sec remains invariant for the two choices.

## 3 Main Result

We next state the main result of the paper after presenting relevant definitions.

**Definition 3.1. Channel Use**: A single channel use over the network is defined as a channel use by a single or a subset of nodes simultaneously accessing the network.

**Definition 3.2. Network Protocol**: A network protocol is a scheme employed over the network that establishes communication between subsets of S-D pairs in single or many channel uses.

**Definition 3.3.** A rate $\Gamma_{sd}$ bits/sec between a S-D pair is said to be achievable, under the network protocol, if there exists a sequence of $(|\mathcal{M}|, b) = (2^{b\Gamma_{sd}}, b)$ codes such that the maximal probability of error, $P_{error}$ in decoding goes to zero as $b \to \infty$. The maximum is taken over all the messages. In terms of bits, the number of bits per message that are reliably communicated is $\Gamma_{sd}b$.

**Definition 3.4. Average network sum rate** is defined as the sum of bits per message over all S-D pairs that are reliably communicated divided by the total number of channel uses employed by the network protocol. Precisely,

$$\sum_{(s,d)} C_n^{s,d} = \frac{\sum_{(s,d)} \Gamma_{s,d} b}{\text{no. of channel uses}}$$

**Encoding:** At each node $v \in V$, we construct a $(2^{b\Gamma}, b)$ Gaussian codebook, i.e. each message $m \in \{1, 2, ..., 2^{b\Gamma}\}$ is assigned a codeword $X_m(1), ..., X_m(b)$ where $X_m(i)$ is i.i.d $\mathcal{N}(0, 1)$. To send the message $m$ the node transmits the codeword $X_m(1), .., X_m(b)$. The distribution over the message set is uniform. The message sets and the codebooks are independent for different nodes. We reveal the codebooks to the respective destinations.

*Remark* 3.1. The reason for constructing equal rate codebooks at each source node is due to the fact that the network protocol employed in the paper is symmetric with respect to all the source destination pairs. This leads to same rate per source destination pair.

We have the following main result.

**Theorem 3.1. *Main Result*:** *Consider the network of S-D pairs and the communication model 2 with attenuation factor $\alpha > 2$. Then there exists a network protocol such that the average network sum rate scales as*

$$\sum_{s,d} C_n^{sd} \geq c' n^{2/3}$$

*and is achievable for some constant $c' > 0$ independent of $n$. This implies that,*

$$\rho(n) \geq \frac{c''}{n^{1/3}}$$

*for some $c'' > 0$ independent of $n$.*

The rest of the paper is focused on the proof of the main result. The proof of the main result rests on construction of a network protocol. The next section presents preliminary notation. We will then provide an outline of the protocol operation in the following section.

## 4 Notation

Denote by $V = \{v_1, v_2, .., v_{n/2}\}$ the set of transmit nodes and by $W = \{w_1, w_2, ..., w_{n/2}\}$ the set of corresponding receive nodes. To simplify the notation we will use $v$ to denote an element of $V$ and $w$ to denote an element of $W$. Let $M = \lfloor n^{1/3} \rfloor$. On the receive side we partition the set of receive nodes into $M/2$ receive clusters each of dimension $M \times M$ and containing $M^2$ nodes. Denote the set of receive clusters by $\mathcal{R} = \{R_1, R_2, .., R_{M/2}\}$. Each receive cluster is further partitioned into $M$ sub clusters each of size $\sqrt{M} \times \sqrt{M}$. Note that there are $M$ sub clusters in each receive cluster $R$. Let $\mathcal{S} = \{S_1, S_2, .., S_{\frac{M^2}{2}}\}$ denote the set of receive sub clusters in all the receive clusters. Finally, we index the sub-clusters within each receive cluster $R$ as $s_1(R), .., s_M(R)$.

**Definition 4.1.** The distance between two receive clusters is defined by

$$\rho(R_i, R_j) = \min_{w \in R_i, w' \in R_j} d(w, w')$$

where $d(.,.)$ is the Euclidean distance.

The distance between the sub clusters is defined in a similar way.

**Definition 4.2.** A receive cluster group, $G_R$ is a subset of $\mathcal{R}$ such that, $\forall R_i, R_j \in G_R, i \neq j$, $\rho(R_i, R_j) \geq 2\sqrt{2}M$. We index different receive cluster groups as $G_R(q)$, $q = 1, 2, ...$

**Definition 4.3.** A sub cluster group $G_S$ is a subset of $\mathcal{S}$ such that, $\forall S_i, S_j \in G_S, i \neq j$, $\rho(S_i, S_j) \geq 2\sqrt{2M}$. We index different receive cluster groups as $G_S(q)$, $q = 1, 2, ...$

The idea is to operate different cluster groups in different channel uses of the channel. Due to the distance separating different clusters in a cluster group, the inter-cluster interference can be minimized and all the different clusters in a cluster group can operate simultaneously. We have the following lemma that bounds the number of receive cluster groups and sub cluster groups.

**Lemma 4.1.** *There exists (a) $c_0 < 19$ disjoint receive cluster groups $G_R(q), q = 1, 2, .., c_0$ such that $\bigcup_q G_R(q) = \mathcal{R}$, and (c) $c_0 < 19$ disjoint sub cluster groups $G_S(q), q = 1, 2, .., c_0$ such that $\bigcup_q G_S(q) = \mathcal{S}$.*

*Proof.* See appendix □

Let $V(R)$ denote the set of transmit nodes that have a destination in the receive cluster $R$. Note that due to one to one correspondence of S-D pairs (transmit/receive nodes) we have, $|V(R)| = M^2$. Further define

$$\tilde{V}(R) = \left\{ v_i \in \tilde{V}(R) : \rho(v_i, R) \geq k_0 M \right\}$$

Basically the subset $\tilde{V}(R) \subset V(R)$ are the set of transmit nodes that are sufficiently far away from the receive cluster. Although, this impacts the sum rate as some sources corresponding to each receive cluster $R$ are not admitted the effect on the ratio $\rho(n)$ is insignificant as we establish later in Section 6.4.

Let the sub cluster containing the node $w_i$ be denoted by $S(w_i)$. Similarly let the receive cluster containing the node $w_i$ be denoted by $R(w_i)$.

**Definition 4.4.** For each sub cluster group $G_S(q)$, define the collection of nodes,

$$\mathcal{W}(G_S(q)) = \{w : S(w_i) \neq S(w_j) \,\forall i \neq j, \, S(w_i) \in G_S(q)\}$$

i.e. $\mathcal{W}(G_S(q))$ selects one node from each of the sub-clusters belonging to a sub-cluster group $G_S(q)$. Note that there are $M$ such collections for each $G_S(q)$. Label them by $\mathcal{W}_p(G_S)(q), p = 1, 2, .., M$.

## 5 Outline of Network Protocol

The network protocol is illustrated in figure 1. The set $\tilde{V}(R)$ simultaneously transmit messages. The $M$ sub clusters in receive cluster $R$ serve as $M$ MIMO relays that help in simultaneously decoding messages from $M$ of the different sources in $\tilde{V}(R)$. These MIMO relays are then time shared for decoding of other sources in $\tilde{V}(R)$. The rate gains over the network comes from operating $M$ such MIMO relay systems (i.e. corresponding to $M$ receive clusters) simultaneously over the network. To realize MIMO relaying a local collaborative scheme is employed within the receive sub clusters.

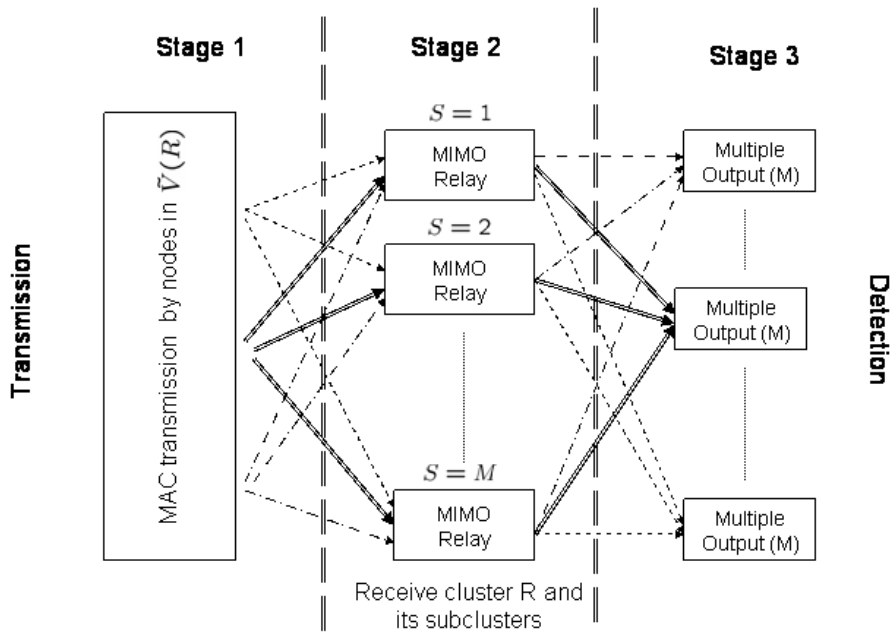

Figure 1: There are essentially three phases in the Network protocol. (1) Multiple access channel (MAC) transmission: All the nodes in $\tilde{V}(R)$ access the channel at the same time. (2) Local information exchange for collaboration (3) Coherent MIMO detection. Stages (2) and (3) constitute MIMO relaying, which is an amplify and forward scheme by a group of M-antenna systems which employ coherent strategy to detect M messages (streams) at M nodes simultaneously. The MIMO relays are *shared* among the nodes in $\tilde{V}(R)$ successively. The rate gains come from operating many such systems simultaneously over the network.

**Phase 1: Transmission**: Consider a receive cluster $R$. Each node in $\tilde{V}(R)$ picks a message from its respective codebook and transmits the corresponding codeword in $b$ channel uses. Nodes in $\tilde{V}(R)$ transmit at the same time. This is repeated for each receive cluster $R \in \mathcal{R}$.

Note that transmission by nodes in $\tilde{V}(R)$ and reception by nodes in $R$ virtually forms a MAC system. Transmission from each node suffers interference from $\leq M^2$ other transmissions. Indeed if the decoding is centralized on the receiver side, then each of the nodes in $\tilde{V}(R)$ can transmit messages successfully to its destination in $R$ at a fixed positive rate, [18, 19] by successive cancellation or by coherent decoupled detection, [20]. But due to decentralization we need to exchange received observations via a noisy wireless fading channel in order to do further processing.

**Phase 2: Local information exchange**: Observations are exchanged (over a noisy channel) via amplify and forward scheme among the nodes in sub clusters belonging to a sub cluster group, $G_S(q)$. To do this, nodes in $\mathcal{W}_p(G_S(q))$ simply broadcast the data to neighboring nodes. Sufficient separation among the sub cluster in sub cluster group ensures that the inter cluster interference from simultaneously operating nodes is limited. This is repeated for all $p$ (for a fixed $q$) and for all $q$, i.e.

for all the $M$ collections $\mathcal{W}_p(G_S(q)), p = 1, 2, .., M$ and for all cluster groups $G_S(q), q = 1, 2, .., c_0$.

**Phase 3: Coherent MIMO detection**: After the exchange is completed, a coherent MIMO detection strategy is employed to detect the transmitted messages in each receive cluster $R$. This coherent MIMO detection is again an amplify and forward strategy whereby conditioned on knowledge of the channel gains [1] nodes preprocess the exchanged data coherently to detect messages at respective nodes. Receive clusters belonging to a receive cluster group do so simultaneously. As will be shown later sufficient separation between the clusters in the cluster group limits inter cluster interference during the coherent detection process.

In the following sections we will prove the main result by: (1) Computing the number of channel uses in each phase of the network protocol; (2) Derive upper bounds for probability of error in decoding messages at the desired destinations.

The first task is straightforward. For the second task our idea is to compute a lower bound for the mutual information achieved between the S-D pairs under the network protocol and show that it is non-vanishing and independent of $n$. Then by exploiting the joint asymptotic equipartition property (AEP), [21], over the memoryless ergodic channel established via the network protocol we argue that the probability of error $P_e$ in decoding goes to zero as $b \to \infty$ for each S-D pair for all rates $\Gamma$ below the achievable mutual information.

## 6 Proof of the main result

### 6.1 Transmission phase

Fix a receive cluster $R$. Each node in $\tilde{V}(R)$ simultaneously transmits a codeword of length $b$ in $b$ channel uses. This is repeated for all the receive clusters in $\mathcal{R}$. We have the following lemma for the number of channel uses in transmission phase.

**Lemma 6.1.** *The total number of channel uses taken to complete the transmission phase is $\frac{bM}{2}$.*

*Proof.* See appendix. □

The actual achievable rate, $\Gamma$ will be quantified based on the mutual information achieved by the particular decoding scheme employed as a part of the network protocol. To this end reader is asked to assume that $\Gamma > 0$. It is important to note that the other receive clusters are inactive while the transmission for a particular receive cluster is going on. This means that after the transmission phase is completed each receive node $w_i$ receives $b$ observations corresponding to all the sources in $\tilde{V}(R)$.

The nodes $\tilde{V}(R)$ are located at different distances from the cluster $R$. Also the distance from a single transmitter in $\tilde{V}(R)$ to different nodes in $R$ is different. This leads to non-uniform receive power and non-uniform channel gains. In order to make the average receive power similar at the nodes in $R$, nodes in $\tilde{V}(R)$ employ a "Power control scheme".

**Lemma 6.2.** *For the above setup there exists a "Power Control scheme" employed by nodes in $\tilde{V}(R)$ such that the receive vector $\mathbf{Y}_R$, for each channel use, at nodes in receive cluster $R$ is given by,*

$$\mathbf{Y}_R = \sqrt{SNR_0}\left[\mathbf{H}_R \circ (\mathbf{1} + \Delta)\right]\mathbf{X}_R + \mathbf{N}_1 \qquad (4)$$

---
[1]One can show similar results for the case when only an estimate of the channel is available. However in this paper we will not deal with such a case

where **1** is a matrix of all ones and $\Delta \in \mathbb{R}^{M^2 \times |\tilde{V}(R)|}, : |\Delta|_{i,j} \leq \frac{\alpha}{k_0}$ for an appropriate (admissible) choice of symbol power for each transmit node and with noise variance $\mathbf{N}_1 \sim \mathcal{N}(0, I_M)$. The $\circ$ denotes the Hadamard product of matrices.

*Proof.* See appendix. □

## 6.2 Local Information exchange

**Exchange Process**: All receive nodes $w_i \in \mathcal{W}_p(G_S(q))$ simultaneously broadcast (amplify and forward) the received observations to nodes in the sub cluster $S(w_i)$. For each receive node, it takes $b$ channel uses to transmit all the received observations. This is repeated for all the collections $p = 1, 2, .., M$ and all sub cluster groups $q = 1, .., c_0$. Thus we have,

**Lemma 6.3.** *The number of channel uses required for local exchange of received observations is $c_0 M b$ channel uses.*

*Proof.* See appendix. □

**Amplify and Forward for exchange**: Let $Y_{w_i}(k)$ denote the observation at the node $w_i \in \mathcal{W}_p(G_S(q))$ in the corresponding $k^{th}$ channel use during the transmission phase. Note that $\mathbf{E}(Y_{w_i}(k))^2 \approx M^2$. Since each node is constrained to transmit a symbol of variance $\leq 1$, the node $w_i$ scales $Y_{w_i}(k)$ by a factor $\xi_1$ and then forwards $\frac{1}{\xi_1} Y_{w_i}(k)$ to all its sub-cluster nodes in $S(w_i)$.

Now each node $w_j \in S(w_i) : i \neq j$ receives the transmitted symbol with a channel fading gain $f_{w_i, w_j}$ and with an attenuation factor $(\frac{d_{max}}{d_{w_i, w_j}})^{\alpha/2}$ according to the communication model. Upon reception node $w_j$ scales the received observation by a factor $(\frac{d_{w_i, w_j}}{d_{max}})^{\alpha/2}$. Therefore, at the end of the exchange process, corresponding to $b^{th}$ channel use, each node $w_j$ has a vector of the form

$$\mathbf{Z}(w_j)(k) = \frac{1}{\xi_1} \mathbf{F}(w_j) \mathbf{Y}_{S(w_j)}(k) + \mathbf{N}_2(k) + \mathbf{N}_{exchange}(k) \qquad (5)$$

Note that here $\mathbf{N}_2$ is the additive receiver noise at node $w_j$ and $\mathbf{N}_{exchange}$ is the interference from other nodes in $\mathcal{W}_p(G_S(q))$ that are accessing the channel simultaneously. The matrix $\mathbf{F}(w_j)$ is the matrix (diagonal) of fading channel gains from all the nodes $w_i : i \neq j$ in $S(w_j)$ to node $w_j$.

Since the processing of the received observations corresponding to different channel uses in the transmission phase is the same, we will drop the indexing on $k$ in the following. We characterize the nature of the exchange noise via the following lemma.

**Lemma 6.4.** *Fix $S = 1$. Then at a node $w \in S = 1$, the variance of the noise process $\mathbf{N}_{exchange}$ obeys,*

$$Var(\mathbf{N}_{exchange}) \leq Var \left[ \sqrt{SNR_0} \sum_{k=1}^{\sqrt{\frac{n}{8M}}-1} \frac{\sqrt{2M}^{\alpha/2}}{(2k\sqrt{2M})^{\alpha/2}} \sum_{S \in G_S, S \neq 1} \mathcal{I}_k(S) \frac{1}{\xi_1}(\mathbf{1} + \Delta) \circ \mathbf{H}_R(S) \mathbf{X}_R \right] \qquad (6)$$

where $\mathcal{I}_k(S)$ is the indicator function whose value is one if the sub-cluster $S$ lies in the annulus centered at sub-cluster $S = 1$, and with radii, $\left(2k\sqrt{2M}, 2(k+1)\sqrt{2M}\right)$. The term $(\mathbf{1}+\Delta) \circ \mathbf{H}_R(S) \in \mathbb{C}^{M \times |\tilde{V}(R)|}$ is the matrix of channel gains from the $|\tilde{V}(R)|$ transmitting nodes to the sub cluster $S \in R$ and $\mathbf{X}_R = \mathbf{X} \in \mathbb{R}^{|\tilde{V}(R)| \times 1}$ corresponding to the transmission for receive cluster $R$.

To illustrate Equation 6, first note that $\mathbf{N}_{exchange}$ is the interference arising from all those sub-clusters that are simultaneously exchanging observations, which are all the sub-clusters belonging to a sub-cluster group. Notice also that the variance of the fading gain from simultaneously transmitting nodes in other sub-clusters is unity (after suitable scaling). The expression inside the inner summation follows from the model (c.f. lemma 6.2), for received observations at the nodes in sub-cluster $S$. The factor of $\frac{1}{\xi_1}$ is due to the scaling of the observations before noisy forwarding.

*Proof.* See Appendix. □

### 6.3 Coherent MIMO Detection

Consider a receive cluster group $G_R(q)$. There are $M$ sub clusters in each $R \in G_R(q)$, which are indexed as $s_1(R),..,s_M(R)$. Coherent detection procedure consists of the following operation: Fix a sub cluster, $s_1(R)$ (say) in each receive cluster $R \in G_R(q)$. The other $M-1$ sub clusters in each receive cluster $R \in G_R(q)$, coherently process (the strategy is outlined in the following section) the exchanged observations for decoding $M$ messages simultaneously at $M$ nodes in sub-cluster $s_1(R)$. This means that in each receive cluster $R \in G_R(q)$, $M$ messages are simultaneously decoded in $b$ channel uses through coherent MIMO detection. This operation is done for all sub clusters $s_1(R),...,s_M(R)$ for all $R \in G_R(q)$ and subsequently for all groups $q = 1,2,..,c_0$. We have the following lemma.

**Lemma 6.5.** *The number of channel uses to complete the MIMO detection process for all the nodes in each receive cluster $R$ takes no more than $c_0 bM$ channel uses.*

*Proof.* See appendix. □

We now need to do the following. (1) Present the coherent detection strategy, and (2) Prove that using the network protocol, decoding of $M$ messages at $M$ nodes is possible with $P_e \to 0$ as $b \to \infty$. Indeed proving the second half will also give us the bound on the overall rate $\Gamma$ of the code book at each node.

#### 6.3.1 Coherent MIMO Detection strategy

Without loss of generality we focus on one receive cluster say $R$. Due to symmetry of the network protocol the following analysis holds for all the receive clusters. The detection strategy depends on the fading that occurs in each of the phases, namely fading during transmission ($\mathbf{H}_R$), fading during the exchange $\mathbf{F}(w) \ \forall w \in R$, and fading from the nodes in $M-1$ sub clusters to the nodes in the sub cluster at which the messages are decoded. Let the matrix of fading gains from the nodes in sub cluster $S$ in receive cluster $R$ to the nodes in subcluster at which coherent detection is taking place be denoted by $\mathbf{Q}_S^1$.

The nature of the coherent detection strategy employed is as follows. Each node $w \in R$ transmits a symbol $Z_1(w) = \mathbf{U}(w)\mathbf{Z}(w)$. The row vector $\mathbf{U}(w)$ depends on the channels $\mathbf{H}_R$ and $\mathbf{Q}_{S(w)}^1$. We will now specify the vectors $\mathbf{U}(w)$. To simplify the analysis and gain insight into the nature of the coherent detection process we identify two cases.

Case 1. Let us assume for the moment that matrices $\mathbf{F}(w)$ is an identity matrix, i.e. there is no fading during the exchange process. Then each node has observation vector of the form,

$$\mathbf{Z}(w) = \frac{1}{\xi_1}\mathbf{Y}_{S(w)} + \mathbf{N}_{exchange} + \mathbf{N}_2$$

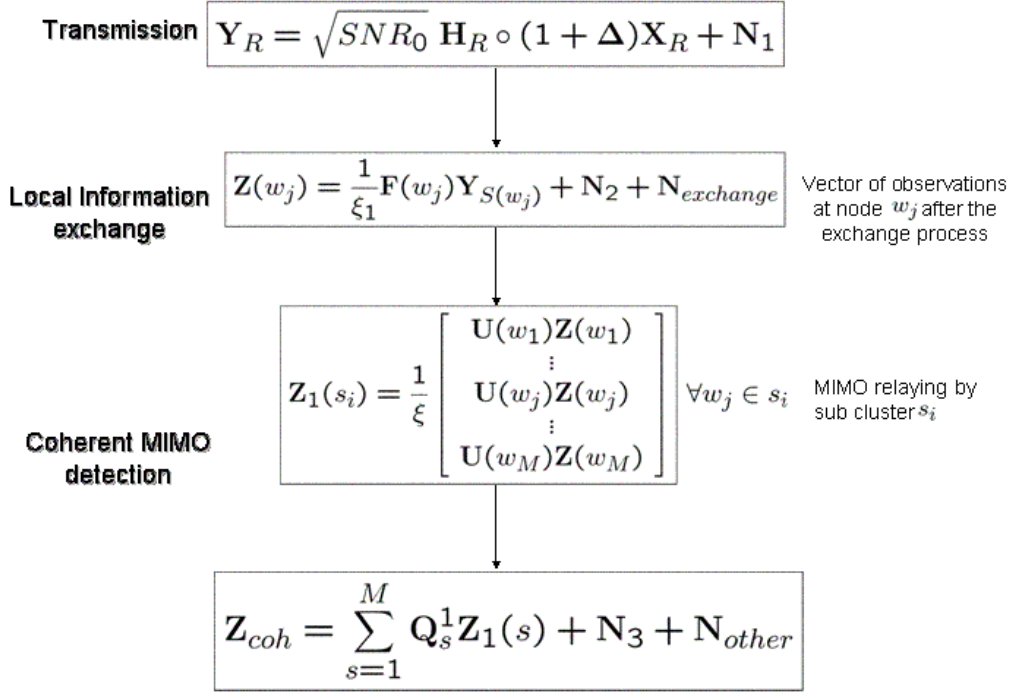

Figure 2: Figure depicting the equations in various stages of the scheme. The set of equations are for detecting $M$ messages that are meant for sub cluster $s_1$ in the receive cluster $R$. The equations represent the processing for each channel use corresponding to receive cluster $R$ in the transmission phase. The processing for other sub clusters is the same and is done over successive channel uses. The receive clusters belonging to a cluster group do so simultaneously, which leads to rate gains over the network.

In this case the processing vector $\mathbf{U}(w)$ is the $w^{th}$ row of the matrix $\frac{1}{\xi}[\mathbf{Q}^1_{S(w)}]^*[\mathbf{H}^{S(w)}_1]^*$. The matrix $\mathbf{H}^{S(w)}_1 \in \mathbb{C}^{M \times M}$ is the matrix of fading gains from the first $M$ transmit nodes in $\tilde{V}(R)$ to the nodes in $S(w)$. The scaling by $\frac{1}{\xi}$ is to ensure that the transmitted symbol is of variance $\leq 1$.

Case 2. When the fading is present during the exchange process, after normalization by the gain factor each node has an observation vector of the form,

$$\mathbf{Z}(w) = \frac{1}{\xi_1}\mathbf{F}(w)\mathbf{Y}_{S(w)} + \mathbf{N}_{exchange} + \mathbf{N}_2$$

For coherent MIMO detection node $w$ first performs the following linear operation: $\mathbf{F}^*(w)\mathbf{Z}(w)$. Then node $w$ linearly processes the resulting vector by the corresponding row of the matrix $\frac{1}{\xi}[\mathbf{Q}^1_{S(w)}]^*[\mathbf{H}^{S(w)}_1]^*$. Again the factor of $\frac{1}{\xi}$ is to ensure that each node transmits a symbol of variance $\leq 1$.

To implement the above coherent detection strategy, each node $w \in R$ needs to know the channel $\mathbf{H}^{S(w)}_1$ and only needs to know the vector of fading coefficients corresponding to the first row of the matrix $\mathbf{Q}^1_{S(w)}$, which is transmit channel state information (CSI) from node $w$ to $M$ nodes in sub cluster at which the coherent detection is taking place. Again we emphasize that it is possible to show that only an estimate of the channel states suffices. However, we do not pursue this here.

### 6.3.2 Achievable rate per source-destination pair

In this section we will calculate a lower bound on the achievable rate for any source-destination pair served by the network protocol. Without loss of generality we will focus on the detection process in one receive cluster. Also since within a receive cluster the processing is same for all the sub clusters we will focus on detection at one particular receive sub cluster. Due to symmetry of the protocol, the results will hold for all receive clusters and sub clusters.

Fix $s_1(R)$ for $R = 1$ as the sub cluster at which the messages are decoded and assume that the first $M$ elements of $\mathbf{X}_R$ denoted by $\mathbf{X}_1$, (say), are meant for the nodes in sub cluster $s_1(R)$. Recall that the $i^{th}$ element, $\mathbf{X}_1(i)$ of vector $\mathbf{X}_1$ is the first symbol of the codeword corresponding to $i^{th}$ node in subcluster $s_1(R)$. Let $\mathbf{Q}_s^1(R)$ denote the channel from a sub cluster $s \in \{s_1(R), ..., s_M(R)\}$ to sub cluster $s_1(R)$ in receive cluster $R$. Since the receive cluster is clear we drop the indexing on $R$ in subsequent analysis.

Under the coherent detection strategy employed, let the vector of received observations under such a strategy at the $M$ nodes in $s_1$ be denoted by $\mathbf{Z}_{coh}$. Then we have,

$$\boxed{\mathbf{Z}_{coh} = \sum_{s=1}^{M} \mathbf{Q}_s^1 \mathbf{Z}_1(s) + \mathbf{N}_3 + \mathbf{N}_{other}} \tag{7}$$

where $\mathbf{Z}_1(s)$ is the vector of symbols transmitted from the nodes in subcluster $s$ in receive cluster $R$. To this end, we have the following lemma.

**Lemma 6.6.** *For the two cases, viz., (1) No fading during the exchange process and (2) fading during the exchange process, along with the respective processing described above the mutual information between the $i^{th}$ component of vector $\mathbf{Z}_{coh}$ and the $i^{th}$ component of vector $\mathbf{X}_1$ is given by,*

$$I(\mathbf{X}_1(i); \mathbf{Z}_{coh}(i)) \geq \frac{1}{2} \log(1 + \beta_2) := \Gamma_2 \tag{8}$$

*for some $\beta_2 > 0$.*

In order to prove the lemma 6.6 we will utilize the following result of [22].

**Lemma 6.7.** *Let*
$$Y = AX_G + W = \hat{A}X_G + (A - \hat{A})X_G + W$$

*where $X_G \sim \mathcal{N}(0, E_s)$ is independent of the pair $(A, \hat{A})$. Assume that $\mathbf{E}(A|\hat{A}) = \hat{A}$ and that conditioned on $(A, \hat{A})$ the random variables $X_G$ and $W$ are independent. Then*

$$I(X_G; Y, \hat{A}) \geq \log\left(1 + \frac{\hat{A}^2 E_s}{\mathbf{E}[|W|^2|\hat{A}] + E_s \mathbf{E}[|A - \hat{A}|^2|\hat{A}]}\right)$$

*Furthermore, the above inequality also holds if conditioned on $\hat{A}$ the random variables $X_G$ and $W + (A - \hat{A})X_G$ are uncorrelated.*

We use the above lemma in the following manner. Observe that the interference process $\mathbf{N}_{exchange}$, $\mathbf{N}_{other}$ and the signal of interest are not independent. This significantly complicates the computation of mutual information. Furthermore, determining the expected value of the conditional mutual information ((conditioned on channel knowledge) over all the random channel realizations, is extremely complicated. The above lemma holds so long as we decompose the received information, $\mathbf{Z}_{coh}$, into a signal term and a residual noise term that are uncorrelated (not necessarily independent). Furthermore, if the decomposition is such that the multiplying factor corresponding to the signal of interest is deterministic then the expectations over random channel realizations of the mutual information can be bounded from below by mutual information of a AWGN channel with the noise power equal to the residual noise power.

**Probability of error in decoding:** Consider the $b$ length codeword $X_v(1), ..., X_v(b)$ transmitted by node $v$ during the transmission phase. Let the received sequence of observations at the respective destination corresponding to transmission by node $v$ under the above coherent detection strategy be given by $Z_{coh,v}(1), ..., Z_{coh,v}(b)$. Under the network protocol employed the effective channel from node $v$ to the destination node is ergodic and memoryless, i.e.,

$$p\left(Z_{coh,v}(1), ..., Z_{coh,v}(b) | X_v(1), ..., X_v(b)\right) = \prod_{k=1}^{b} p\left(Z_{coh,v}(k) | X_v(k)\right) = p^b\left(Z_{coh,v} | X_v\right)$$

This implies that the random variables $X_v$ and $Z_{coh,v}$ satisfy the joint AEP, [21]. Then under the uniform distribution over the codewords, it follows from, [21] that the maximal probability of error when using jointly typical decoding is upper bounded by,

$$P_e \leq 2^{-b(\Gamma - I(X_v; Z_{coh,v}))}$$

From lemma 6.6, $I(X_v : Z_{coh,v}) \geq \frac{1}{2} \log(1 + \beta_2)$ for all $v$. Thus, if the rate of the codebook at each node is $\Gamma < \frac{1}{2}\log(1 + \beta_2)$ then probability of error $P_e \to 0$ as $b \to \infty$. In particular, for any $\epsilon > 0$, a rate of $\Gamma_2 - \epsilon$ is achievable for any source-destination pair. We can choose $\epsilon$ to be as small as desired so that $\Gamma = \Gamma_2 - \epsilon > 0$.

## 6.4 Average network sum rate

We combine the above results to determine the achievable network sum rate. For this since we allow only the nodes in $V'(R)$ to transmit, we need to calculate how many sources are served in the scheme. For this first recall that,

$$\tilde{V}(R) = \left\{v_i \in \tilde{V}(R) : \rho(v_i, R) \geq k_0 M\right\}$$

The total number of sources that are served is given by, $\sum_R \tilde{V}(R)$. Remove a strip of size $k_0 M \times \sqrt{n}$ from the middle, and do not allow the transmissions from these sources. Then the total number of sources that are served is equal to $\frac{1}{2}(n - k_0 n^{5/6}) = \frac{n}{2}(1 - \frac{k_0}{n^{1/6}})$. For $k_0 << n^{1/6}$ the total number of sources that are served $\geq c_1 n$ for $c_1 > 0$.

We have the following proposition for the total number of channel uses employed by the network protocol.

**Proposition 6.8.** *The total number of channel uses employed by the network protocol is $(2c_0 + 1)Mb$.*

*Proof.* The proof follows from lemmas 6.1, 6.3 and 6.5. □

Each source-destination pair that is served by the network protocol gets $b$ channel uses to transmit a message in every $(2c_0 + 1)Mb$ channel uses by the network protocol. Thus effectively the rate of message transmission per source destination pair drops by a factor of $\frac{1}{(2c_0+1)M}$ due to the channel sharing in the network protocol. This is adequately accounted for in the calculation of the average network sum rate below.

**Average network sum rate** : From above the average network sum rate achievable by the network protocol obeys,

$$\sum_{s,d} C_n^{sd} \geq \frac{c_1 n \Gamma_2 b}{(M + c_0 M + c_0 M)b} \geq c' n^{2/3}$$

for some $c' > 0$.

**The metric $\rho(n)$** : Note that $\sum_{s,d} C_0^{sd} = \frac{n}{4} \log(1 + SNR_0)$. Hence from the above it follows that,

$$\rho(n) \geq \frac{c''}{n^{1/3}}$$

for some $c'' > 0$. This proves the main result.

# 7 Conclusions and future research directions

In this paper we provided lower bounds to the capacity of wireless networks under assumptions of fixed SNR in the network. Fixed SNR makes the network capable of cancelling interference under rich scattering environment. Distributed collaborative schemes proposed in this paper uses this capability leading to increase in throughput over the traditional multi-hop schemes. Typically fixed SNR, high scattering wireless networks can be formed in urban areas where there are lots of scatterers present and the geographical expanse is limited. In information theoretic terms, degrees of freedom in a scattering environment depends on the number of independent channels that can be supported by the environment. These independent channels can be shared over subsets of source destination pairs to allow for more transmissions while cancelling the multi-user interference via collaboration. Our paper presents one such network protocol in this direction. One important point to note is that due to decentralization, there is an inherent tradeoff between the extent of collaboration required to cancel multi-user interference and number of users that can simultaneously be supported.

# 8 Appendix

## 8.1 Proof of lemma 4.1

*Proof.* It is sufficient to prove the result for the set of receive clusters. The arguments are same for the case of sub clusters. To this end identify the set of receive clusters with a vertex set $V$ of a graph, i.e., $V = \mathcal{R}$. The edge set $E$ of the graph is given as follows. Connect two vertices in $V$ if the distance between the vertices is $\leq 2\sqrt{2}M$. Then we get an almost regular graph where the degree of each node is bounded by $c_0 \leq 2.\pi.2\sqrt{2} + 1 < 19$. From the graph coloring lemma [26], it follows that one can color the vertices of the graph by using no more than $c_0$ colors such that no two adjacent vertices are of the same color. Define $G_R(1)$ to be the set of vertices which are of color 1 and so on. Thus there are $c_0$ disjoint sets $G_R \subset \mathcal{R}$ with the required properties. $\square$

## 8.2 Proof of lemma 6.1

*Proof.* There are $\frac{M}{2}$ receive clusters $R$. Corresponding to each receive cluster $R$ nodes $\tilde{V}(R)$ transmit massages for the destinations in cluster $R$ in $b$ channel uses. Over all the receive clusters it takes $\frac{bM}{2}$ channel uses to complete the transmission phase of the protocol. $\square$

## 8.3 Proof of lemma 6.2

Consider a transmit node $v \in \tilde{V}(R)$. Consider the center of the receive cluster $R$ and call it $R_c$. Let node $v$ transmit with power $(\frac{d(v,R_c)}{d_{\max}})^\alpha$. Then at node $w$ in receive cluster $R$ the gain according to the model of equation (3) is given by,

$$h_{vw}\sqrt{SNR_0}(\frac{d_{\max}}{d(v,w)})^\alpha (\frac{d(v,c)}{d_{\max}})^\alpha = h_{vw}\sqrt{SNR_0}(\frac{d(v,c)}{d(v,w)})^\alpha$$

Note that,

$$\frac{d(v,c)}{d(v,c)+2M} \leq \frac{d(v,c)}{d(v,w)} \leq \frac{d(v,c)}{d(v,c)-2M}$$

Since $d(v,c)$ is greater than $k_0 M$, this implies that,

$$\frac{1}{1+2/k_0} \leq \frac{d(v,c)}{d(v,w)} \leq \frac{1}{1-2/k_0}$$

For sufficiently large value of $k_0$ the gain

$$h_{vw}\sqrt{SNR_0}(\frac{d(v,c)}{d(v,w)})^{\alpha/2} \sim h_{vw}\sqrt{SNR_0}(1 \pm \frac{\alpha}{k_0})$$

This proves the lemma.

## 8.4 Proof of lemma 6.3

*Proof.* Since each node receives $b$ observations of interest, it takes $b$ channel uses per node in the collection to forward its observations. Over $M$ such collections per sub cluster group it takes $Mb$ channel uses. For all the $c_0$ cluster groups it takes $c_0 Mb$ channel uses. $\square$

## 8.5 Proof of lemma 6.4

*Proof.* Fix a sub-cluster say $S = 1$. The gain at node $w' \in (S=1)$ from nodes in other sub-clusters $S \neq 1$, such that $\mathcal{I}_k(S) = 1$ is upper bounded by $\frac{d_{\max}^{\alpha/2}}{(2k\sqrt{2M})^{\alpha/2}}$. This follows from the scale invariant communication model.

Since the nodes in sub cluster $S = 1$ normalize the received (during exchange phase) observations by the attenuation gain factor, the final expression follows from the normalization by a factor $(\frac{d_{\max}}{d(w_1,w')})^{\alpha/2}$ and $d(w_1, w') \leq \sqrt{2M}$. $\square$

## 8.6 Proof of lemma 6.5

*Proof.* For each receive cluster $R$ in a receive cluster group $G_R$ it takes $bM$ channel uses to complete the detection process for all the sub clusters. Over all the $c_0$ receive cluster groups it takes $c_0 bM$ channel uses. □

## 8.7 Proof of lemma 6.6

Let us first focus on the case when there is no fading during the exchange process. To simplify the analysis we further modify our strategy. Basically, for a given destination sub cluster $s_1(R)$ in a receive cluster $R$, the sub-clusters in receive cluster $R$, participating in the MIMO detection process are limited to those that are at a set distance of $M/3$ from the destination sub-cluster, i.e.,

$$\tilde{S} = \{\tilde{s} \in \{s_1, .., s_M\} \mid \rho(\tilde{s}, s_1) \geq M/3\}$$

The idea is to ensure that the signals from the different sub-clusters have similar powers. We have the following obvious lemma, which ensures that at least $M/2$ clusters participate in the MIMO detection process.

**Lemma 8.1.** $\left|\tilde{S}\right| \geq M/2$

Without loss of generality assume that only $M/2$ sub clusters are taking part in the MIMO detection process. Also assume that $|\tilde{V}(R)| = M^2$ for the worst case interference. Then Under the coherent detection strategy, the received signal is given by

$$\begin{aligned}\mathbf{Z}_{coh} &= \sum_{\tilde{s}=1}^{M/2} \frac{\sqrt{SNR_0} d_{\max}^{\alpha/2}}{r_{\tilde{s},1}^{\alpha/2}} \frac{1}{\xi}(\mathbf{1}+\Delta_q(\tilde{s})) \circ \mathbf{Q}_{\tilde{s}}^1 [\mathbf{Q}_{\tilde{s}}^1]^* [\mathbf{H}_1^{\tilde{s}}]^* \left[\frac{1}{\xi_1}(\mathbf{J}_{\tilde{s}}\mathbf{X} + \mathbf{N}_2^{\tilde{s}}) + \mathbf{N}_{exchange} + \mathbf{N}_3^{\tilde{s}}\right] \\ &+ \mathbf{N}_{other} + \mathbf{N}_4 \end{aligned} \quad (9)$$

where the factor of $\frac{\sqrt{SNR_0}d_{\max}^{\alpha/2}}{r_{\tilde{s},1}^{\alpha/2}}$ is in accordance with the communication model, the factor $(\mathbf{1}+\Delta_q(\tilde{s}))$ before the matrices $\mathbf{Q}_{\tilde{s}}^1$ is to account for the difference in fading gains due to attenuation, the factors of $\frac{1}{\xi}$ and $\frac{1}{\xi_1}$ are due to the scaling introduced to ensure that the transmitted symbol from each node is of unit power in the coherent MIMO detection and the data exchange process respectively. The matrix $\mathbf{J}_{\tilde{s}} \in \mathbb{R}^{M \times M^2}$ is given by,

$$\mathbf{J}_{\tilde{s}} = (\mathbf{1} + \Delta_h(\tilde{s})) \circ [\mathbf{H}_1^{\tilde{s}}, ..., \mathbf{H}_M^{\tilde{s}}]$$

where $\mathbf{H}_t^{\tilde{s}}, t = 1, .., M$ is the matrix of fading gains from the $\{(t-1)M, ..., tM\}$ transmit nodes in $\tilde{V}(R)$ to nodes in subcluster $\tilde{s}$. It is worthwhile to point out again that $|\tilde{V}(R)| = M^2$ for the worst case interference.

*Remark* 8.1. Note that the elements of the matrix $\Delta_h(\tilde{s})$ and $\Delta_q(\tilde{s})$ are bounded between $\pm \frac{\alpha}{k_0}$ and $\pm \frac{\alpha}{\sqrt{M}}$ respectively. These variations are very small for a fixed $\alpha$ and for sufficiently large value of $k_0$ and for sufficiently large number of nodes, $n$.

*Remark* 8.2. Note that since the distance from the sub clusters to the intended destination is $\geq M/3$ we can fix the distance $r_{\tilde{s},1} = M$. This will indeed not affect the orders of the signal power and that of the interference. Moreover we are interested in the scaling of these quantities rather than their exact calculations.

For sake of notational convenience we will index the sub clusters taking part in the MIMO detection process via $s$ instead of $\tilde{s}$. In what follows $c_i, i \in \mathbb{N}$ are bounded positive constants independent of $n$.

Let,

$$\mathbf{X}_R = [\mathbf{X}_1(1), \ldots, \mathbf{X}_1(M), \ldots, \mathbf{X}_M(1), \ldots, \mathbf{X}_M(M)]^T \tag{10}$$

In the above representation $\mathbf{X}_j(i) : i = 1, .., M; j = 1, .., M$ is the symbol from the codeword corresponding to the $i^{th}$ node in the subcluster $s_j$ in the receive cluster $R$.

Without loss of generality, consider the first symbol $\mathbf{X}_1(1)$. Using lemma 6.7 we will prove a lower bound on the mutual information between $\mathbf{X}_1(1)$ and $\mathbf{Z}_{coh}(1)$. First note that the conditions of the lemma are satisfied under the assumption that only the expected channel gain is known to the intended receiver. This makes the interference uncorrelated with the signal of interest. Note that the noises $\mathbf{N}_3, \mathbf{N}_2, \mathbf{N}_4$ are of unit variance in each dimension and are independent zero mean. The noise process $\mathbf{N}_{other}, \mathbf{N}_{exch}$ are uncorrelated with the signal of interest due to uncorrelated fading.

Under processing as shown in equation 9, let the effective channel from $\mathbf{X}_1(1)$ and $\mathbf{Z}_{coh}(1)$ be given by,

$$\mathbf{Z}_{coh}(1) = A\mathbf{X}_1(1) + W$$

where $A$ is the channel gain and where $W$ is the cumulative effect of all the noises and the interferences, i.e., additive noises and multi-user and simultaneous user interferences. Our first task is to find $\mathbf{E}(A)$. Let,

$$\mathbf{Q}_s^1 = \begin{bmatrix} \mathbf{q}_s(1) \\ \mathbf{q}_s(2) \\ \vdots \\ \mathbf{q}_s(M) \end{bmatrix} \quad [\mathbf{H}_1^s]^* = \begin{bmatrix} \mathbf{h}_s^*(1) \\ \mathbf{h}_s^*(2) \\ \vdots \\ \mathbf{h}_s^*(M) \end{bmatrix} \tag{11}$$

We drop the gain factors $\frac{d_{\max}^{\alpha/2}}{r_{s,1}^{\alpha/2}}, \frac{1}{\xi}$ and $\frac{1}{\xi_1}$ in the following calculations as they are common. Then we have the following expression for $A$.

$$A = \sum_{s=1}^{M/2} \sum_{i=1}^{M} [(\mathbf{1} + \delta_1(s,1)) \circ \mathbf{q}_s(1)] \mathbf{q}_s^*(i) \mathbf{h}_s^*(1) [(\mathbf{1} + \delta_2(s,i)) \circ \mathbf{h}_s(i)]$$

where $\delta_1(s,1), \delta_2(s,i)$ are corresponding perturbation column vectors from the matrices $\Delta_q(S)$ and $\Delta_h(S)$. Again for notational simplicity introduce,

$$\tilde{\mathbf{q}}_s(i) = (\mathbf{1} + \delta_1(s,i)) \circ \mathbf{q}_s(i) \ ; \tilde{\mathbf{h}}_s(i) = (\mathbf{1} + \delta_2(s,i)) \circ \mathbf{h}_s(i)$$

Now, in evaluating $\mathbf{E}(A)$, note that for terms with $i \neq 1$ the expected value is zero. Hence, we have

$$\mathbf{E}(A) = \sum_{s=1}^{M/2} \mathbf{E}\left[\tilde{\mathbf{q}}_s(1)\mathbf{q}_s^*(1)\mathbf{h}_s^*(1)\tilde{\mathbf{h}}_s(i)\right] = \sum_{s=1}^{M/2} sum(\mathbf{1} + \delta_1(s,1)).\ sum(\mathbf{1} + \delta_2(s,1)) \geq \frac{cM^3}{2}$$

for some $c > 0$. This is because the perturbations are very small and they don't affect the order. Taking into account the additional gain factors and scaling, the actual gain is of the order

$$\mathbf{E}(A) \geq \frac{\sqrt{SNR_0} d_{\max}^{\alpha/2}}{M^{\alpha/2}} \frac{c}{2\xi_1 \xi} M^3$$

Now we need to evaluate $\mathbf{E}(A - \hat{A})^2$. Since $\hat{A} = \mathbf{E}A$ this expression is equal to $\mathbf{E}(A^2) - \hat{A}^2$. We now calculate $\mathbf{E}(A^2)$. Dropping for the moment the common prefactors (scaling and gains due to attenuation) we have,

$$\mathbf{E}A^2 = \mathbf{E}\left(\sum_{r=1}^{M}\sum_{s=1}^{M}\sum_{i=1}^{M}\sum_{j=1}^{M}[\tilde{\mathbf{q}}_s(1)\mathbf{q}_s^*(i)][\tilde{\mathbf{q}}_r(1)\mathbf{q}_r^*(j)][\mathbf{h}_s^*(1)\tilde{\mathbf{h}}_s(i)][\mathbf{h}_r^*(1)\tilde{\mathbf{h}}_r(j)]\right)$$

Now note that for all $r, s$ and for $i \neq 1, j \neq 1, i \neq j$ the expected value is zero. Also for $r \neq s, i \neq 1, j \neq 1, i = j$ the expected value is zero. So the above expression reduces to

$$\begin{aligned}
\mathbf{E}A^2 &= \mathbf{E}\left(\sum_{r=1}^{M/2}\sum_{s=1}^{M/2}[\tilde{\mathbf{q}}_s(1)\mathbf{q}_s^*(1)][\tilde{\mathbf{q}}_r(1)\mathbf{q}_r^*(1)][\mathbf{h}_s^*(1)\tilde{\mathbf{h}}_s(1)][\mathbf{h}_r^*(1)\tilde{\mathbf{h}}_r(1)]\right) \\
&+ \mathbf{E}\left(\sum_{s=1}^{M}\sum_{i=1,i\neq 1}^{M}||\tilde{\mathbf{q}}_s(1)\mathbf{q}_s^*(i)||^2||\mathbf{h}_s^*(1)\tilde{\mathbf{h}}_s(i)||^2\right) \\
&\leq c_1 \frac{M^6}{4} + c_2 \frac{M^3}{2}(M-2)
\end{aligned} \quad (12)$$

for some $c_1, c_2 > 0$. Incorporating the common gain factor $\frac{SNR_0 d_{\max}^{\alpha}}{M^{\alpha}} \frac{1}{(\xi_1 \xi)^2}$ we have

$$\mathbf{E}(A - \hat{A})^2 \leq \frac{SNR_0 d_{\max}^{\alpha}}{M^{\alpha}} \frac{1}{(\xi_1 \xi)^2} \left(c_3 M^6 + c_2 M^3 (M-2)\right)$$

In order to evaluate the interference power we will evaluate first the multiuser interference arising from the $M^2$ symbols. Then we will show that the other interferences will exhibit the same behavior in power. For simplicity of the exposition we will drop the small deterministic perturbations in the subsequent calculations, since they do not affect the order of the terms. The multi-user interference power is interference power per dimension of the following vector

$$\mathbf{W}_{multiuser} = \sum_{s=1}^{M}\sum_{k=1}^{M}\mathbf{Q}_s^1[\mathbf{Q}_s^1]^*[\mathbf{H}_1^s]^*\mathbf{H}_k^s \mathbf{X}_k$$

We have the following lemma.

**Lemma 8.2.** *The power per dimension in the vector $\mathbf{W}_{multiuser}$ is $\leq c_1 M^6$ for some finite positive $c_1$ independent of $n$.*

*Proof.* Without loss of generality consider the interference power in the first element of the vector. We have,

$$W_{multiuser} = \sum_{s=1}^{M}\sum_{k=1}^{M}\sum_{i=1}^{M}\mathbf{q}_s(1)\mathbf{q}_s^*(i) \sum_{\substack{j=1;\text{ if } k=1, j\neq 1}}^{M} \mathbf{h}_s^*(i)\mathbf{h}_s^k(j)\mathbf{X}_k(j)$$

$$W_{multiuser} = \sum_{k=1}^{M} \sum_{j=1;\text{ if } k=1, j \neq 1}^{M} \left[ \sum_{i=1}^{M} \sum_{s=1}^{M} \mathbf{q}_s(1)\mathbf{q}_s(i)\mathbf{h}_s^*(i)\mathbf{h}_s^k(j)\mathbf{X}_k(j) \right]$$

Since random variables, $\mathbf{X}_k(j)$ for different values of $j, k$ are uncorrelated with each other, we can write.

$$\mathbf{E}(W_{multiuser})^2 = \sum_{k=1}^{M} \sum_{j=1;\text{ if } k=1, j \neq 1}^{M} \mathbf{E}\left[ \sum_{i=1}^{M}\sum_{s=1}^{M}\sum_{l=1}^{M}\sum_{r=1}^{M} \mathbf{q}_s(1)\mathbf{q}_s(i)\mathbf{q}_r(1)\mathbf{q}_r(l)\mathbf{h}_s^*(i)\mathbf{h}_s^k(j)\mathbf{h}_r^*(l)\mathbf{h}_r^k(j) \right]$$

The expectation is non zero if and only if $r = S$ and $i = l$. Hence we have

$$\mathbf{E}(W_{multiuser})^2 = \sum_{k=1}^{M} \sum_{j=1;\text{ if } k=1, j \neq 1}^{M} \mathbf{E}\left[ \sum_{i=1}^{M}\sum_{s=1}^{M} \mathbf{q}_s(1)\mathbf{q}_s(i)\mathbf{q}_s(1)\mathbf{q}_s(i)\mathbf{h}_s^*(i)\mathbf{h}_s^k(j)\mathbf{h}_s^*(i)\mathbf{h}_s^k(j) \right]$$

$$\mathbf{E}(W_{multiuser})^2 = \sum_{k=1}^{M} \sum_{j=1;\text{ if } k=1, j \neq 1}^{M} \mathbf{E}\left[ M^3 + \sum_{i=2}^{M}\sum_{s=1}^{M} \mathbf{q}_s(1)\mathbf{q}_s(i)\mathbf{q}_s(1)\mathbf{q}_s(i)\mathbf{h}_s^*(i)\mathbf{h}_s^k(j)\mathbf{h}_s^*(i)\mathbf{h}_s^k(j) \right]$$

$$\mathbf{E}(W_{multiuser})^2 = \sum_{k=1}^{M} \sum_{j=1;\text{ if } k=1, j \neq 1}^{M} \left[ M^3 + \mathbf{E}\sum_{i=2}^{M}\sum_{s=1}^{M} ||\mathbf{q}_s(1)\mathbf{q}_s(i)||^2 ||\mathbf{h}_s^*(i)\mathbf{h}_s^k(j)||^2 \right]$$

$$\mathbf{E}(W_{multiuser})^2 = \sum_{k=1}^{M} \sum_{j=1;\text{ if } k=1, j \neq 1}^{M} \left[ M^3 + \mathbf{E}\sum_{i=2}^{M}\sum_{s=1}^{M} M^2 \right] \leq c_4 M^6$$

for some $c_4 > 0$. With the gain factor we have

$$\mathbf{E}(W_{multiuser})^2 \leq \frac{SNR_0 d_{\max}^\alpha}{M^\alpha} \frac{1}{(\xi_1 \xi)^2} c_4 M^6$$

$\square$

Now note the following points.

1. The interference vectors inside the summation (c.f. lemma 6.4) $\mathbf{N}_{exchange}$ is of the same form as $\mathbf{W}_{multiuser}$. So the upper bound in the lemma 8.2 of interference power per dimension provides an upper bound to the interference power per dimension from each of the terms inside the summation in $\mathbf{N}_{exchange}$. Specifically,

$$\begin{aligned}
\mathbf{W}_{exch} &= Var(\sum_{s=1}^{M/2} \frac{1}{\xi}(\mathbf{1} + \Delta_q(s)) \circ \mathbf{Q}_s^1 [\mathbf{Q}_S^1]^* [\mathbf{H}_1^s]^* \mathbf{N}_{exchange}) \\
&\leq Var \sum_{s=1}^{M/2} \frac{1}{\xi}(\mathbf{1} + \Delta_q(s)) \circ \mathbf{Q}_s^1 [\mathbf{Q}_S^1]^* [\mathbf{H}_1^s]^* \\
&\quad \cdot \left[ \sqrt{SNR_0} \sum_{k=1}^{\sqrt{\frac{n}{8M}}-1} \frac{\sqrt{2M}^{\alpha/2}}{(2k\sqrt{2M})^{\alpha/2}} \sum_{S \in G_S} \mathcal{I}_k(S) \frac{1}{\xi_1} \mathbf{H}_R(S) \mathbf{X}_R \right]
\end{aligned}$$

In order to calculate the variance of the process on the right hand side, note that for different clusters the channels $\mathbf{H}_R(S)$ are uncorrelated with the signal of interest and for clusters outside the receive cluster under consideration, both $\mathbf{X}_R$ and $\mathbf{H}_R(S)$ are uncorrelated with the signal of interest, i.e. $\mathbf{X}_1(1)$. So the total interference power $\mathbf{W}_{exch}$ due to the exchange process is given by,

$$\sum_{k=1}^{\sqrt{\frac{n}{8M}}-1} \frac{\sqrt{2M}^{\alpha/2}}{(2k\sqrt{2M})^{\alpha/2}} \sum_{S \in G_S} \mathcal{I}_k(S)$$

$$\times \left[ Var \sum_{s=1}^{M/2} \tfrac{1}{\xi}(\mathbf{1} + \mathbf{\Delta}_q(s)) \circ \mathbf{Q}_s^1 [\mathbf{Q}_s^1]^* [\mathbf{H}_1^s]^* \left[ \sqrt{SNR_0} \tfrac{1}{\xi_1}(\mathbf{1}+\mathbf{\Delta}) \circ \mathbf{H}_R(S)\mathbf{X}_R \right] \right]$$

Thus the variance per dimension is bounded by,

$$\mathbf{E}(W_{exch})^2 \leq \frac{SNR_0}{(\xi\,\xi_1)^2} \sum_{k=1}^{\sqrt{\frac{n}{8M}}-1} \frac{1}{(2^\alpha k^\alpha)} \sum_{S \in G_S} \mathcal{I}_k(S) c_4 M^6$$

where we have used the result of lemma 8.2 to upper bound the interference of each of the contributing terms. Now since

$$\sum_{S \in G_S} \mathcal{I}_k(S) \leq 8\pi k$$

which is an upper bound on the number of sub clusters in an annulus with radii $(2k\sqrt{2M}, 2(k+1)\sqrt{2M})$. Then we have,

$$\mathbf{E}(W_{exch})^2 \leq \frac{SNR_0}{(\xi\,\xi_1)^2\, 2^\alpha} \sum_{k=1}^{\sqrt{\frac{n}{8M}}-1} \frac{8\pi}{k^{\alpha-1}} c.M^6 \leq \frac{c_5}{(\xi\,\xi_1)^2\, 2^\alpha} M^6$$

for values of $\alpha > 2$, where $c_5 > 0$ is bounded. With the gain factor we have

$$\mathbf{E}(W_{exch})^2 \leq \frac{SNR_0 d_{\max}^\alpha}{M^\alpha} \frac{c_5}{(\xi\,\xi_1)^2\, 2^\alpha} M^6$$

2. The interference due to noises $\mathbf{N}_2$ is given by

$$\mathbf{W}_{\mathbf{N}_2} = \sum_{s=1}^{M/2} \frac{\sqrt{SNR_0} d_{\max}^{\alpha/2}}{r_{s,1}^{\alpha/2}} \frac{1}{\xi}(\mathbf{1}+\mathbf{\Delta}_q(s)) \circ \mathbf{Q}_s^1 [\mathbf{Q}_s^1]^* [\mathbf{H}_1^s]^* (\frac{1}{\xi_1}\mathbf{N}_2^S)$$

The interference power per dimension in this case is upper bounded by

$$\mathbf{E}(W_{\mathbf{N}_2})^2 \leq \frac{SNR_0 d_{\max}^\alpha}{M^\alpha} \frac{c_6}{(\xi\xi_1)^2} M^4$$

3. The interference due to $\mathbf{N}_3$ is given by

$$\mathbf{W}_{\mathbf{N}_3} = \sum_{s=1}^{M/2} \frac{\sqrt{SNR_0}d_{\max}^{\alpha/2}}{r_{s,1}^{\alpha/2}} \frac{1}{\xi}(\mathbf{1} + \Delta_q(s)) \circ \mathbf{Q}_s^1[\mathbf{Q}_s^1]^*[\mathbf{H}_1^s]^*\mathbf{N}_3^s$$

The interference power per dimension in this case is upper bounded by

$$\mathbf{E}(W_{\mathbf{N}_3})^2 \leq \frac{SNR_0 d_{\max}^\alpha}{M^\alpha} \frac{c_7}{(\xi)^2} M^4$$

4. Note that the interference power per dimension from $\mathbf{N}_4$ is unity.

5. Finally we have to calculate the interference $\mathbf{N}_{other}$ due to simultaneously operating clusters. To do this, note that the total power that a receive cluster is operating with is sum of the powers $\mathbf{E}(W_{multiuser})^2, \mathbf{E}(W_{exch})^2, \mathbf{E}(W_{\mathbf{N}_2})^2, \mathbf{E}(W_{\mathbf{N}_3})^2$ and $\mathbf{E}(A)^2$, without the gain factor due to attenuation. Let this sum power be denoted by $P_{tot}$. For a receive cluster that is at a distance of $2\sqrt{2}M$ from the receive cluster under consideration, the contribution to the interference noise per dimension is upper bounded by, $\frac{SNR_0 d_{\max}^\alpha}{(kM)^\alpha} P_{tot}$. As $k$ goes from 1 to $\frac{d_{\max}}{M}$, the total interference power per dimension in $\mathbf{N}_{other}$ is given by

$$\mathbf{E}(W_{other})^2 \leq \sum_{k=1}^{\frac{d_{\max}}{M}} \frac{SNR_0 d_{\max}^\alpha}{(kM)^\alpha} 8\pi k P_{tot} \leq \frac{SNR_0 d_{\max}^\alpha}{(M)^\alpha} \sum_{k=1}^{\frac{d_{\max}}{M}} \frac{8\pi}{k^{\alpha-1}} P_{tot} \leq \frac{SNR_0 d_{\max}^\alpha}{(M)^\alpha} c_8 P_{tot}$$

for values of $\alpha > 2$ and for some $c_8 > 0$. The above expression follows from the fact that there are no more than $8\pi k$ receive clusters inside the annulus with radii $2\sqrt{2}kM, 2\sqrt{2}(k+1)M$, centered around the receive cluster under consideration.

It remains to calculate $P_{tot}$. To find this first note that the $\xi_1 = c_9 M$ and $\xi = c_{10} M^2$. Thus $P_{tot} \leq c_{11} \frac{M^6}{(\xi\xi_1)^2}$ for some $c_{11} > 0$. Taking into account all the interferences and substituting in the expression for the mutual information we get,

$$I(X_1(1); \mathbf{Z}_{coh}(1)) \geq \log\left(1 + \frac{c^2/4}{c_4 + \frac{c_5}{2^\alpha} + c_6/M^2 + \frac{c_7}{(c_9)^2} + c_8 + c_{11} + \frac{(c_9 c_{10})^2 M^\alpha}{SNR_0 d_{\max}^\alpha}}\right) \geq \log(1 + \beta_2)$$

where $\beta_2 > 0$ is non vanishing and independent of $n$.

## 8.8 Fading during the exchange process

In this section we will prove the lemma when there is fading in the exchange process. As discussed in section 6.3.1, nodes after normalization of the received observation by the attenuation gain factor, multiplies the received observations by the conjugate of corresponding fading gain. In particular the set of observation at a node is given by,

$$\mathbf{Z}(w) = \frac{1}{\xi_1}\mathbf{F}(w)\mathbf{Y}_{S(w)} + \mathbf{N}_{exchange} + \mathbf{N}_3$$

The node does $\mathbf{F}^*(w)\mathbf{Z}(w)$. Then it process the observations by the corresponding row of the matrix $\frac{1}{\xi}[\mathbf{Q}_S^1]^*[\mathbf{H}_1^S]^*$. The symbol $Z_1(k)$ transmitted by the $k$ node in the sub cluster $S$ is given by,

$$\frac{1}{\xi \xi_1} \sum_{m=1}^M \sum_{j=1}^M \sum_{n=1}^M q_{nk}^*(s) h_{jn}^*(1,s) \gamma_j^k \sum_{i=1}^n h_{ji}(m,s) \mathbf{X}_i(m)$$

The vector of symbols transmitted in absence of other noises and interferences, (we will separately account for noises and interferences subsequently) from the nodes in sub cluster $s$ is given by,

$$\mathbf{Z}_1(s) = \begin{bmatrix} \frac{1}{\xi \xi_1} \sum_{m=1}^M \sum_{j=1}^M \sum_{n=1}^M q_{n1}^*(s) h_{jn}^*(1,s) \gamma_j^1 \sum_{i=1}^n h_{ji}(m,s) \mathbf{X}_i(m) \\ \vdots \\ \frac{1}{\xi \xi_1} \sum_{m=1}^M \sum_{j=1}^M \sum_{n=1}^M q_{nk}^*(s) h_{jn}^*(1,s) \gamma_j^k \sum_{i=1}^n h_{ji}(m,s) \mathbf{X}_i(m) \\ \vdots \\ \frac{1}{\xi \xi_1} \sum_{m=1}^M \sum_{j=1}^M \sum_{n=1}^M q_{nM}^*(s) h_{jn}^*(1,s) \gamma_j^M \sum_{i=1}^n h_{ji}(m,s) \mathbf{X}_i(m) \end{bmatrix} \quad (13)$$

where $\gamma_j^k$ is the $(j,j)$ th element of the matrix $\mathbf{F}^*(w_k)\mathbf{F}(w_k)$. $h_{ji}(m,s)$ is the $(j,i)$th element of the matrix $\mathbf{H}_m^s$ and $q_{nk}$ is the $(n,k)$th element of the matrix $\mathbf{Q}_s^1$. The vector of received observations at sub-cluster 1 is then given by,

$$\mathbf{Z}_{coh} = \sum_{s=1}^{M/2} \frac{\sqrt{SNR_0} d_{max}^{\alpha/2}}{r_{s,1}^{\alpha/2}} \mathbf{Q}_S^1 \mathbf{Z}_1(s) + \frac{1}{\xi}[\mathbf{Q}_s^1]^*[\mathbf{H}_1^s]^* \left(\frac{1}{\xi_1}\mathbf{N}_2^s + \mathbf{N}_{exchange} + \mathbf{N}_3^s\right) + \mathbf{N}_{other} + \mathbf{N}_4$$

*Remark* 8.3. Note that in the above expression we have dropped the perturbation factors $(\mathbf{1}+\Delta_q(S))$ and $(\mathbf{1} + \Delta_h(S))$ as these factors do not affect the order of the powers in the signal and in the interferences, as we have seen in the previous case. The main aim is to see whether the fading during the exchange process affects the result.

In the above expression the noises $\mathbf{N}_3$ and $\mathbf{N}_2$ are still zero mean noise vectors with i.i.d components. The variance does not change with the change in processing, i.e. operation $\mathbf{F}^*$. Similarly the variance of per dimension of the noise vector $\mathbf{N}_{exchange}$ due to extra processing with the matrix $\mathbf{F}^*$. Now we need to calculate the signal gain and the interference powers. Dropping the common gain factors as before, the expected signal gain in $X_1(1)$ is given by

$$\mathbf{E}(A) = \mathbf{E} \sum_{s=1}^{M/2} \sum_{k=1}^M \sum_{j=1}^M \sum_{n=1}^M q_{1k}(s) q_{nk}^*(s) h_{jn}^*(1,s) \gamma_j^k h_{j1}(1,s)$$

Note that the expectation is zero if $n \neq 1$. For $n = 1$ we have,

$$\mathbf{E}(A) = \mathbf{E} \sum_{s=1}^{M/2} \sum_{k=1}^M \sum_{j=1}^M q_{1k}(s) q_{1k}^*(s) h_{j1}^*(1,s) \gamma_j^k h_{j1}(1,s)$$

Taking the expectation inside and noting that $\mathbf{E}(q_{1k}q_{1k}^*(s)) = 1$ and $\mathbf{E}h_{j1}^*(1,s)\gamma_j^k h_{j1}(1,s) = \mathbf{E}(h_{j1}^*(1,s)h_{j1}(1,s))\mathbf{E}(\gamma_j^k) = 1$ we have, $\mathbf{E}(A) = M^3/2$. Taking into account the effect of perturbations in the effective channel gains due to difference in perturbations we have $\mathbf{E}(A) \geq cM^3/2$ for some $c > 0$. Incorporating the gain factors we get $\mathbf{E}(A) \geq c\frac{\sqrt{SNR_0}d_{max}^{\alpha/2}}{M^{\alpha/2}\xi\xi_1}M^3/2$

Now we will calculate $\mathbf{E}(A^2)$. This is given by

$$\mathbf{E}A^2 = \sum_{s,s',k,k',n,n',j,j'=1}^{M} q_{1k}(s)q_{nk}^*(s)q_{1k'}(s')q_{n'k'}^*(s')h_{jn}^*(1,s)\gamma_j^k h_{j1}(1,s)h_{j'n'}^*(1,s')\gamma_{j'}^{k'} h_{j'1}(1,s')$$

$$\begin{aligned}\mathbf{E}A^2 &= \mathbf{E}\sum_{s,s',k,k',j,j'=1}^{M} q_{1k}(s)q_{1k}^*(s)q_{1k'}(s')q_{1k'}^*(s')h_{j1}^*(1,s)\gamma_j^k h_{j1}(1,s)h_{j'1}^*(1,s')\gamma_{j'}^{k'} h_{j'1}(1,s') \\ &+ \mathbf{E}\sum_{s,s',k,k',j,j'=1}^{M}\sum_{n,n'=2}^{M} q_{1k}(s)q_{nk}^*(S)q_{1k'}(s')q_{n'k'}^*(s')h_{jn}^*(1,s)\gamma_j^k h_{j1}(1,s)h_{j'n'}^*(1,s')\gamma_{j'}^{k'} h_{j'1}(1,s')\end{aligned}$$

The first summation in the above expression evaluates to $M^6$. In the second summation note that if $n \neq n'$ then the expected value is zero. Since $n \neq 1, n' \neq 1$ in the second summation the expected value is again zero if $k \neq k'$ and $j \neq j'$ and $s \neq s'$. Thus the second summation is equal to

$$= \mathbf{E}\sum_{s,k,j=1}^{M}\sum_{n=2}^{M} q_{1k}(s)q_{nk}^*(s)q_{1k}(s)q_{nk}^*(S)h_{jn}^*(1,s)\gamma_j^k h_{j1}(1,s)h_{jn}^*(1,s)\gamma_j^k h_{j1}(1,s) = M^3(M-2)$$

Thus we have after taking into account the effect of perturbation terms we have for some $c_1 > 0$,

$$\mathbf{E}(A-\hat{A})^2 \leq c_1 M^6$$

Again, incorporating the common gain factors we have,

$$\mathbf{E}(A-\hat{A})^2 \leq c_1 \frac{SNR_0 d_{\max}^\alpha}{M^\alpha(\xi\xi_1)^2} M^6$$

In order to find the order of the multi-interferences in this case, we will first evaluate the interference due to one symbol in $\mathbf{X}_R$ and then add over all the symbols. To this end consider the interference $W(\mathbf{X}_i(m))$, for a fixed $m$ such that if $i = 1$ then $m \neq 1$. Then,

$$W(\mathbf{X}_i(m)) = \mathbf{E}\left[\sum_{s=1}^{M}\sum_{k=1}^{M}\sum_{j=1}^{M}\sum_{n=1}^{M} q_{1k}(s)q_{nk}^*(s)h_{jn}^*(1,s)\gamma_j^k h_{ji}(m,s)X_i(m)\right]^2$$

$$= \mathbf{E}\sum_{s,s',k,k',j,j',n,n'=1}^{M} q_{1k}(s)q_{nk}^*(s)q_{1k'}(s')q_{n'k'}^*(s')h_{jn}^*(1,s)\gamma_j^k h_{ji}(m,s)h_{j'n'}^*(1,s')\gamma_{j'}^{k'} h_{j'i}(m,s')$$

Now in the above expression if $n \neq n'$ and $j \neq j'$ then expected value is zero.

$$W(\mathbf{X}_i(m)) = \mathbf{E}\sum_{s,s',k,k',j,n=1}^{M} q_{1k}(s)q_{nk}^*(s)q_{1k'}(s')q_{nk'}^*(s')h_{jn}^*(1,s)\gamma_j^k h_{ji}(m,s)h_{jn}^*(1,s')\gamma_j^{k'} h_{ji}(m,s')$$

Now note that of $s \neq s'$ then the expected value is zero. Thus we have

$$W(\mathbf{X}_i(m)) = \mathbf{E} \sum_{s,k,k',j,n=1}^{M} q_{1k}(s)q_{nk}^*(s)q_{1k'}(s)q_{nk'}^*(s)h_{jn}^*(1,s)\gamma_j^k h_{ji}(m,s)h_{jn}^*(1,s)\gamma_j^{k'} h_{ji}(m,s)$$

$$= \mathbf{E} \sum_{s,k,k',j,n=1}^{M} q_{1k}(s)q_{nk}^*(s)q_{1k'}(s)q_{nk'}^*(s)||h_{jn}^*(1,s)||^2 \gamma_j^k ||h_{ji}(m,s)||^2 \gamma_j^{k'}$$

Now if $k \neq k'$ then the expected value is zero. Thus we have,

$$W(\mathbf{X}_i(m)) = \mathbf{E} \sum_{s,k,j,n=1}^{M} q_{1k}(s)q_{nk}^*(s)q_{1k}(s)q_{nk}^*(s)||h_{jn}^*(1,s)||^2 \gamma_j^k ||h_{ji}(m,s)||^2 \gamma_j^k$$

$$= \sum_{s,k,j,n=1}^{M} \mathbf{E}||q_{1k}(s)||^2 \mathbf{E}||q_{nk}^*(s)||^2 \mathbf{E}||h_{jn}^*(1,s)||^2 \mathbf{E}||\gamma_j^k||^2 \mathbf{E}||h_{ji}(m,s)||^2 = M^4$$

Thus the interference from one symbol i.e. $\mathbf{X}_i(m)$ is of the order $M^4$. After taking into account the effect of perturbations in the channel gains due to difference in attenuation we have $W(\mathbf{X}_i(m)) \leq c_2 M^4$ for some $c_2 > 0$. There are $M^2 - 1$ such symbols in $\mathbf{X}_R$ for $i = 1,..,M; m = 1,..,M$ such that for $m = 1, i \neq 1$. So the total multi-user interference is bounded by $c_3 M^6$ per dimension. Incorporating the gain factors the total multi-user interference noise is upper bounded by $c_3 \frac{SNR_0 d_{max}^\alpha}{M^\alpha (\xi\xi_1)^2} M^6$ per dimension.

Using the above result and following similar arguments as in the previous case with no fading we can bound the effect of interference from $\mathbf{N}_{exchange}$ by $c_4 \frac{SNR_0 d_{max}^\alpha}{2^\alpha M^\alpha (\xi\xi_1)^2} M^6$ per dimension. The contribution from the noises $\mathbf{N}_2$ and $\mathbf{N}_3$ is also upper bounded by $c_5 \frac{SNR_0 d_{max}^\alpha}{M^\alpha (\xi\xi_1)^2} M^4$ and $c_6 \frac{SNR_0 d_{max}^\alpha}{M^\alpha (\xi)^2} M^4$ respectively. Following similar arguments as in the previous case the total interference from simultaneously operating receive clusters is upper bounded by,

$$W_{other} \leq c_7 \frac{d_{max}^\alpha}{M^\alpha} M^6$$

Again the normalizing factors $\xi$ and $\xi_1$ are bounded by $c_9 M$ and $c_{10} M^2$ respectively. Plugging the values of these expressions into the formula for the mutual information we have,

$$I(X_1(1), \mathbf{Z}_{coh}(1)) \geq \log \left( 1 + \frac{c^2/4}{c_1 + c_3 + \frac{c_4}{2^\alpha} + c_5/M^2 + \frac{c_6}{(c_9)^2} + c_7 + \frac{(c_9 c_{10})^2 M^\alpha}{SNR_0 d_{max}^\alpha}} \right) \geq \log(1 + \beta_2')$$

for some $\beta_2' > 0$ and independent of $n$.